\documentclass[9pt,twocolumn,twoside]{opticajnl}
\journal{opticajournal}
\setboolean{shortarticle}{false}
\setboolean{displaycopyright}{false}
\dates{}
\doi{}
\nolinenumbers
\title{Experimental demonstration of phase-conjugated twin-wave coherent reception over 100 km with injection locking assisted phase stabilization}
\author[1]{Zhenlin Zhao}
\author[2]{Xingwen Yi}
\author[1,*]{Dawei Wang}
\affil[1]{Guangdong Provincial Key Laboratory of Optoelectronic Information Processing Chips and Systems, School of Electronics and Information Technology, Sun Yat-sen University, Guangzhou, 510006, China}
\affil[2]{The DSP Centre of Excellence, School of Computer Science and Engineering, Bangor University, Dean Street, LL57 1UT Bangor, United Kingdom}
\affil[*]{\href{mailto:wangdw9@mail.sysu.edu.cn}{wangdw9@mail.sysu.edu.cn}}

\begin{abstract}
Recovering complex optical signals with high sensitivity is central to coherent communication and sensing.
Beyond conventional homodyne and heterodyne reception, phase-conjugate twin-wave encoding and phase-sensitive joint detection offer an alternative strategy with enhanced sensitivity.
Extending this approach to long-distance transmission of high-order-modulated signals is challenging because optical phase coherence must be maintained at the receiver.
Here we experimentally demonstrate injection-locking-assisted phase-conjugated twin-wave reception of 200-MBaud signals with modulation formats up to 256-QAM over 100~km of optical fiber.
Optical injection locking regenerates the local oscillator (LO) from a reference transmitted through the 16~dB-loss link, while pilot-assisted optical feedback stabilizes the relative signal--LO phase, enabling coherent combining of the phase-conjugated twin waves at a single balanced photodetector.
At equal received signal-photon number per quadrature in each data-carrying optical mode, the measured PCTW signal-to-noise ratio (SNR) exceeds the ideal single-mode homodyne standard quantum limit (SQL) benchmark.
The SNR gains over the measured homodyne and conventional heterodyne results are about 3~dB and 6~dB, respectively.
Constellation and bit-error-rate measurements show that this gain improves high-order data recovery relative to heterodyne reception.
An ideal Gaussian-noise model predicts a photon information efficiency of 1.54~bits per received signal photon for uniform 16-QAM.
These results establish an injection-locking-assisted coherent receiver architecture that implements joint detection of phase-conjugated twin waves over 100~km of fiber with enhanced sensitivity, providing a practical route for photon-efficient coherent communication.
\end{abstract}

\begin{document}
\raggedbottom
\maketitle

\section{Introduction}

Coherent optical receivers based on optical homodyne and heterodyne detection play critical roles in optical communications~\cite{Winzer2018FiberNetworks,Kikuchi2016Fundamentals,Ip2008Coherent}, continuous-variable quantum key distribution~\cite{Zhang2024CVQKDReview,Grosshans2003CoherentQKD}, precision optical spectroscopy~\cite{Coddington2016DualComb}, and optical quantum-state tomography~\cite{Tiunov2020Tomography,Smithey1993Tomography}.
For weak received signals, recovering complex modulation with high sensitivity requires controlling both detection noise and the optical phase reference.
The information recovered per signal photon also depends on the signal modulation and detection mechanism~\cite{Banaszek2020QuantumLimits,Lee2016GaussianBenchmark}.
Ideal homodyne detection measures only one quadrature at a time but can reach the standard quantum limit (SQL) in signal-to-noise ratio (SNR).
Heterodyne detection recovers both quadratures but suffers from a 3~dB SNR reduction due to additional vacuum noise from the image band~\cite{Kikuchi2016Fundamentals,Shapiro2009QuantumTheory}.
Beyond these two conventional methods, phase-conjugate encoding and joint detection offer an alternative to conventional single-mode reception.
It encodes a complex symbol in two optical modes with conjugate amplitudes and recovers it through joint measurement of the mode pair.
In the spectral implementation considered here, phase-conjugated twin-wave (PCTW) encoding uses two sidebands symmetrically placed about the local oscillator (LO), allowing a phase-sensitive detector to perform the joint measurement.
PCTW coherent reception recovers both quadratures while eliminating the 3~dB SNR degradation because the image band contains a coherent signal copy rather than pure noise.
This strategy is particularly useful when image-band noise pollution critically affects system performance: one can use the image band, if optically available, for better receiving sensitivity.

Early studies have recognized the advantages of this particular information encoding and reception from the perspective of quantum-state estimation.
In 2001, Cerf and Iblisdir showed that a conjugate coherent-state pair permits more precise displacement estimation than two identical coherent-state copies~\cite{CerfIblisdir2001}.
In 2007, Niset and co-workers experimentally demonstrated higher reconstruction fidelity with joint measurements for a Gaussian ensemble of phase-conjugate coherent-state pairs~\cite{Niset2007}.
Barbosa and co-workers developed a detailed two-sideband description of spectral detection, identifying the joint optical observables contained in the electronic photocurrent~\cite{Barbosa2013}.
Sch\"afermeier and Andersen subsequently analyzed the communication performance of phase-conjugate encoding with joint Gaussian detection, quantifying the achievable mutual information and investigating further enhancements enabled by entanglement~\cite{SchafermeierAndersen2018}.

PCTWs have also been employed in classical fiber communication to mitigate fiber nonlinearity, with the signal and its conjugate replica transmitted on orthogonal polarization states~\cite{Liu2013PCTW}.
In that implementation, the two optical fields are recovered through separate I/Q measurements using a polarization- and phase-diverse coherent receiver, then combined digitally to suppress nonlinear distortion~\cite{Liu2013PCTW}.
This post-detection combination differs from a phase-sensitive joint optical measurement of the conjugate modes.
In the ideal shot-noise-limited regime, each separately measured mode retains the conventional 3~dB per-quadrature SNR penalty relative to homodyne detection.

Related implementations of phase-conjugate encoding and joint detection also exist across several areas of applied optics.
In 2015, Li and co-workers demonstrated phase-sensitive balanced detection of symmetric optical sidebands using a single local oscillator (LO) derived from the same laser~\cite{Li2015Sideband}.
Their experiment verified the noise-performance difference between single- and double-sideband reception using continuous-wave tones, providing a direct measurement-level precedent, but without demonstrating long-distance transmission of complex data modulation.
Conjugated-pair photonic homodyne down-conversion has also been demonstrated in radio-over-fiber (RoF) transmission~\cite{Sakamoto2018ConjugatedRoF}, alongside simplified coherent receivers for high-optical-budget analog radio links~\cite{Milovancev2021Simplified}.
Nonlinear phase-sensitive amplification (PSA) offers another route to sensitive coherent reception~\cite{Tong2011Ultrasensitive,Kakarla2020OnePhoton,Larsson2024SingleWave}, which has been demonstrated in long-haul transmission~\cite{Olsson2018LongHaulPSA} and, more recently, in high-capacity transmission over field-deployed fiber~\cite{Chen2025FieldPSA}.
In these classical PSA-assisted communication demonstrations, phase-sensitive optical processing is followed by conventional coherent detection of the amplified signal.
We have previously demonstrated phase-conjugated subcarrier encoding with phase-sensitive joint detection in a 20 km continuous-variable quantum key distribution system~\cite{Su2023PCS}, and more recently, the same system combined with PSA~\cite{Liao2025PSACVQKD}.

However, direct application of phase-conjugate encoding and joint detection to high-order modulation over long-distance fiber transmission in a classical communication framework remains largely unexplored.
Here, we experimentally demonstrate PCTW-based information encoding with modulation formats up to 256-QAM and phase-sensitive joint detection using a single balanced photodetector over 100~km of optical fiber.
We address the challenge of distributing a sufficiently coherent optical phase reference to the remote receiver to coherently combine the conjugate modes, despite propagation loss and fiber-induced phase fluctuations, by using optical injection locking~\cite{LiuSlavik2020OIL,Kakarla2018OIL,Kakarla2020OnePhoton,Su2022OIL} for LO regeneration, and pilot-assisted optical feedback for active phase-locking of signal and LO.
Optical injection locking has enabled high-order QAM transmission~\cite{Kasai2015InjectionLocked256QAM,Wang2018InjectionLockedQAM} and phase-coherent optical-frequency transfer~\cite{Kim2015CoherentTransfer}.
Combining injection locking with electronic feedback has also supported stable homodyne reception and optical carrier recovery~\cite{Fice2011OIPLLReceiver,Liu2015HomodyneOFDM}.
At equal received signal-photon number per quadrature in each data-carrying optical mode, the measured PCTW SNR exceeds the ideal single-mode homodyne SQL benchmark.
The SNR gains over the measured homodyne and conventional heterodyne results are about 3~dB and 6~dB, respectively.
At one total received signal photon per conjugate-mode pair, the model predicts a photon information efficiency (PIE) of 1.54~bits per photon for uniform 16-QAM.
For comparison, a single-mode heterodyne reception gives 0.99~bits per photon under the same conditions.
Together, these results demonstrate the practical implementation of conjugate-mode joint reception for long-distance high-order coherent communication and establish an experimental basis for developing photon-efficient optical links.

\begin{figure*}[!t]
\centering
\includegraphics[width=0.80\textwidth]{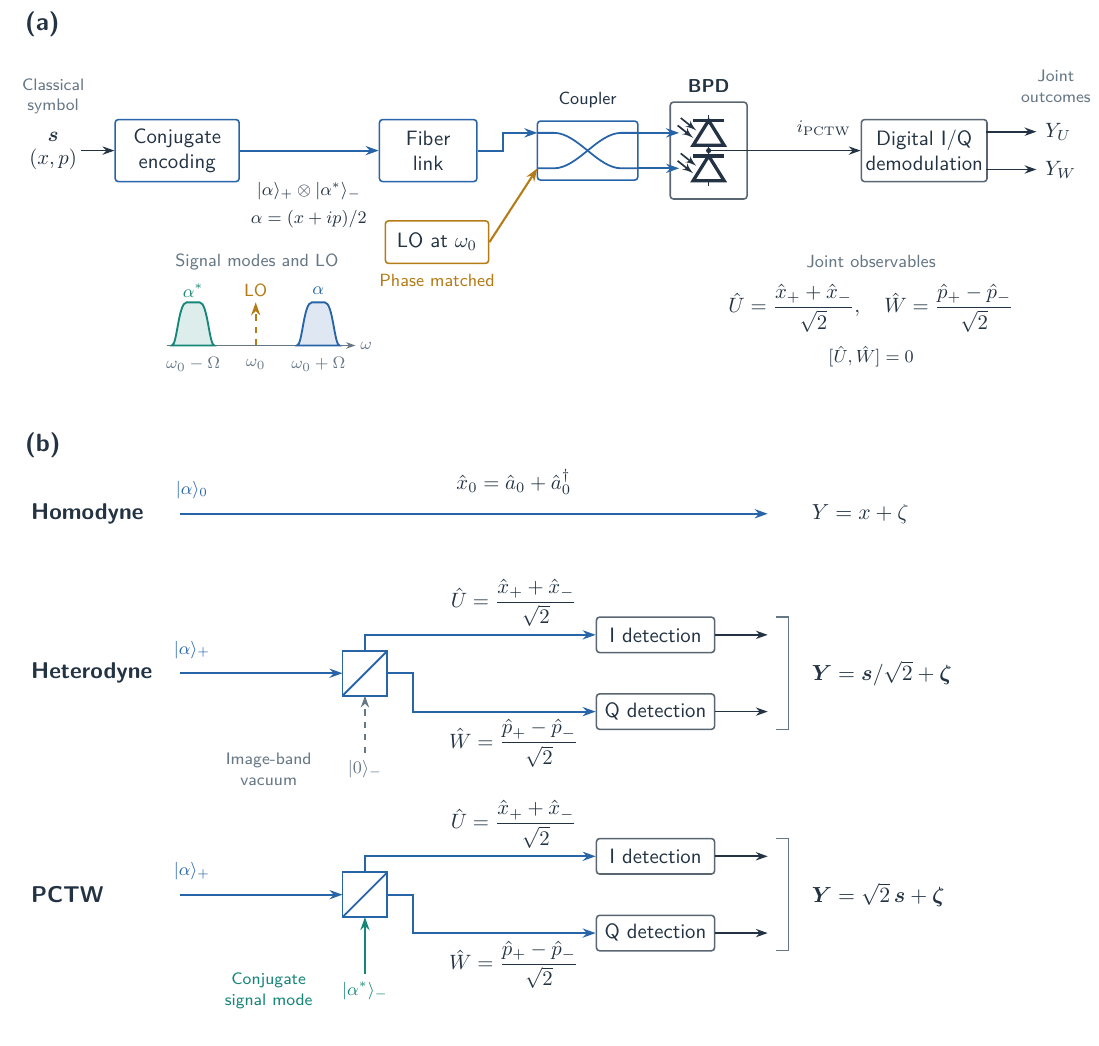}
\caption{
\textbf{PCTW encoding, joint reception, and equivalent measurement models.}
(a) A classical symbol $\boldsymbol{s}$ is encoded in two coherent modes with conjugate amplitudes, centered at $\omega_0\pm\Omega$.
After fiber transmission, the conjugate modes are mixed with a phase-matched local oscillator (LO) at $\omega_0$ and detected using a balanced photodetector (BPD).
Digital in-phase and quadrature (I/Q) demodulation of the photocurrent recovers the commuting joint observables $\hat U$ and $\hat W$.
(b) Equivalent optical-mode descriptions of homodyne, heterodyne, and PCTW detection.
Homodyne detection measures one selected quadrature $x$ or $p$, whereas heterodyne and PCTW detection recover the symbol $\boldsymbol{s}$.
The image-band input is vacuum in conventional heterodyne detection and a conjugate coherent-state signal in PCTW detection.
The comparison assumes equal variance per modulated quadrature of each signal mode.
}
\label{fig:pctw_encoding_and_detection}
\end{figure*}

\begin{figure*}[!t]
  \centering
  \includegraphics[width=1.0\textwidth]{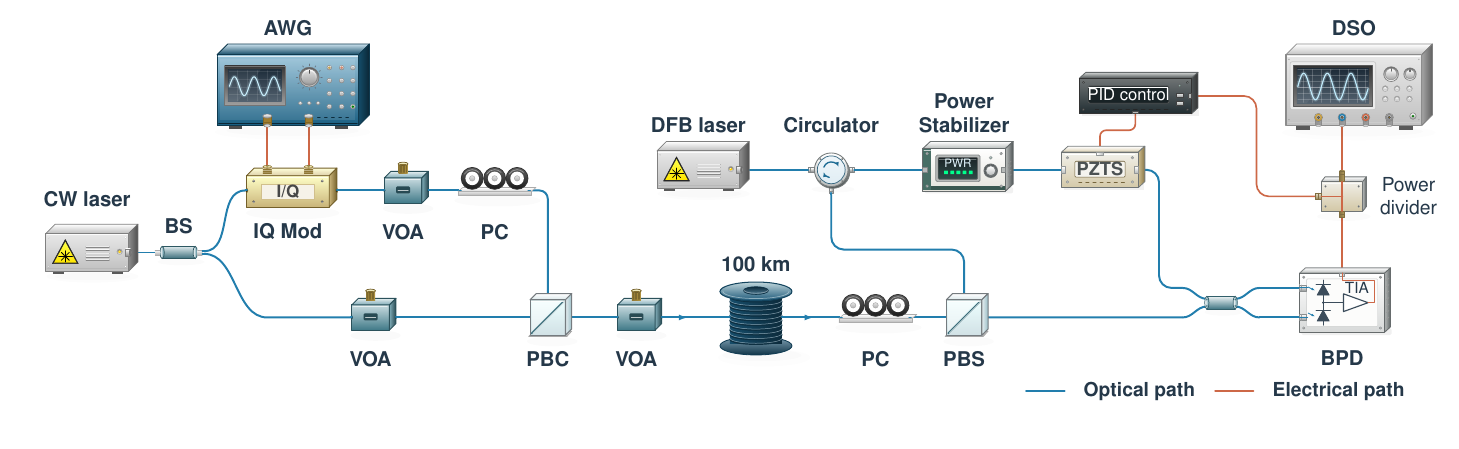}
  \caption{
  \textbf{Experimental setup for PCTW transmission over \(100~\mathrm{km}\) with injection-locking-assisted phase-sensitive reception.}
  The signal and reference are polarization-multiplexed and transmitted   At the receiver, the reference injection-locks a slave laser to regenerate the LO, while auxiliary-pilot feedback stabilizes the signal--LO relative phase.
  Blue and red lines denote optical and electrical paths, respectively.
}
  \label{figure2}
\end{figure*}

\section{Operation Principles}

Figure~\ref{fig:pctw_encoding_and_detection} illustrates the schematic of PCTW encoding and joint reception, together with the equivalent measurement models of homodyne, heterodyne, and PCTW detection.
PCTW encoding represents a classical symbol $\boldsymbol s=(x,p)^{\mathsf T}$ by two coherent modes with conjugate amplitudes, centered at $\omega_0\pm\Omega$:
\begin{equation}
|\Psi_{\boldsymbol s}\rangle
=|\alpha\rangle_+\otimes|\alpha^*\rangle_-,
\qquad
\alpha=\frac{x+ip}{2}.
\label{eq:pctw-alphabet}
\end{equation}
We use $\hat a_j=(\hat x_j+i\hat p_j)/2$, with unit vacuum variance in each quadrature~\cite{Weedbrook2012Gaussian}.
For a zero-mean complex alphabet with covariance $\operatorname{Cov}(\boldsymbol s)=V_M I_2$, each mode carries $V_M/2$ signal photons on average.
When the PCTW and LO are phase-matched, PCTW joint detection measures the commuting observables~\cite{CerfIblisdir2001,Niset2007,Arnhem2019}
\begin{equation}
\begin{gathered}
\hat U=\frac{\hat x_++\hat x_-}{\sqrt2},\qquad
\hat W=\frac{\hat p_+-\hat p_-}{\sqrt2},\\
[\hat U,\hat W]=0.
\end{gathered}
\label{eq:pctw-bell}
\end{equation}
With a strong single-frequency LO at $\omega_0$, the complex radio-frequency photocurrent from balanced detection is proportional to~\cite{Barbosa2013,Li2015Sideband}
\begin{equation}
\hat a_++\hat a_-^\dagger
=\frac{\hat U+i\hat W}{\sqrt2}.
\label{eq:pctw-spectral-current}
\end{equation}
Electronic I/Q demodulation therefore recovers both joint observables from a single balanced-detector output.
For matched conjugate modes and ideal detection, the normalized measurement outcomes satisfy
\begin{equation}
\boldsymbol Y=\sqrt2\,\boldsymbol s+\boldsymbol\zeta,
\qquad
\boldsymbol\zeta\sim\mathcal N(\boldsymbol0,I_2).
\label{eq:pctw-awgn}
\end{equation}
The conjugate signal amplitudes thus add coherently, while each joint observable retains unit vacuum-noise variance.

\subsection{Detection SNR}
\label{subsec:pctw-snr}

At equal modulation variance $V_M$ in each modulated quadrature of every signal mode, the ideal SNRs of homodyne, heterodyne, and PCTW are given by
\begin{equation}
\begin{aligned}
\mathrm{SNR}_{\mathrm{hom}}&=V_M,\\
\mathrm{SNR}_{\mathrm{het}}&=\frac{V_M}{2},\\
\mathrm{SNR}_{\mathrm{PCTW}}&=2V_M.
\end{aligned}
\label{eq:pctw-per-mode-snr}
\end{equation}
Under this per-mode normalization, ideal PCTW reception provides a 3~dB SNR gain over the single-mode homodyne SQL benchmark and a 6~dB gain over conventional heterodyne reception~\cite{Li2015Sideband}.
PCTW occupies two signal modes, so this comparison assigns twice the total signal-photon number to PCTW as to either single-mode reference.

The comparison changes when the total received signal-photon number is fixed.
For the same balanced two-quadrature alphabet, the corresponding SNRs at a common total signal-photon budget $N$ are given by
\begin{equation}
\begin{aligned}
\mathrm{SNR}_{\mathrm{hom}}&=2N,\\
\mathrm{SNR}_{\mathrm{het}}&=N,\\
\mathrm{SNR}_{\mathrm{PCTW}}&=2N.
\end{aligned}
\label{eq:pctw-total-photon-snr}
\end{equation}
Here, homodyne detection recovers only one encoded quadrature, while PCTW recovers both with the same per-quadrature SNR.
Relative to heterodyne reception, the ideal PCTW advantage is therefore 3~dB at equal total signal photons.

\subsection{Mutual information and PIE}
\label{subsec:pctw-information}

For a circular Gaussian alphabet, the ideal mutual information under the same total signal-photon budget can be expressed as
\begin{equation}
\begin{aligned}
I_{\mathrm{hom}}(N)&=\frac12\log_2(1+2N),\\
I_{\mathrm{het}}(N)&=\log_2(1+N),\\
I_{\mathrm{PCTW}}(N)&=\log_2(1+2N).
\end{aligned}
\label{eq:pctw-gaussian-mi}
\end{equation}
These rates are expressed in bits per encoded symbol, with one PCTW symbol occupying two conjugate optical modes~\cite{SchafermeierAndersen2018}.
The homodyne expression retains the same two-quadrature alphabet but observes only one coordinate.
If all photons are instead allocated to real-valued Gaussian modulation, its rate becomes $\tfrac12\log_2(1+4N)$~\cite{TakeokaGuha2014}; two independently real-modulated homodyne modes can match the PCTW rate at the same total photon budget.

Define the photon information efficiency as
\begin{equation}
\mathrm{PIE}(N)=\frac{I(N)}{N}.
\label{eq:pctw-finite-pie}
\end{equation}
For Gaussian modulation, the PCTW predicts $\mathrm{PIE}(N)=\log_2(1+2N)/N$, which approaches $2/\ln2\simeq2.89$~bits per photon as $N\to0$.
For uniform square m-QAM, evaluating mutual information requires numerical estimation~\cite{Alvarado2018AIR}.
At $N=1$, the predicted PIE is approximately 1.543 and 1.554~bits per received signal photon for uniform 16-QAM and 256-QAM, respectively.

\subsection{Phase-sensitive reception}
\label{subsec:pctw-phase}

Preserving the optical phase relation between the conjugate modes and LO is critical for PCTW joint detection.
Let $\phi_+$ and $\phi_-$ denote the phase shifts of the two sidebands, and let $\theta$ denote the LO phase.
Define
\begin{equation}
\delta=\frac{\phi_++\phi_-}{2}-\theta,
\qquad
\psi=\frac{\phi_+-\phi_-}{2}.
\label{eq:pctw-phase-definitions}
\end{equation}
For matched received amplitudes, the joint measurement becomes
\begin{equation}
\begin{aligned}
\boldsymbol Y&=\sqrt2\,(\cos\delta)\,R(\psi)\boldsymbol s+\boldsymbol\zeta,\\
R(\psi)&=\begin{pmatrix}
\cos\psi&-\sin\psi\\
\sin\psi&\cos\psi
\end{pmatrix}.
\end{aligned}
\label{eq:pctw-phase-output}
\end{equation}
The differential phase $\psi$ rotates the recovered constellation, whereas the mean-phase mismatch $\delta$ reduces the coherent-combining amplitude.
After compensating the output rotation, the per-quadrature SNR is
\begin{equation}
\mathrm{SNR}_{\mathrm{PCTW}}(\delta)
=
2N\cos^2\delta.
\label{eq:pctw-phase-snr}
\end{equation}
Digital phase correction can remove $\psi$, but cannot restore the SNR lost due to signal-LO phase mismatch.
Therefore, optical phase stabilization is essential to realize PCTW joint detection for reliable high-order data recovery after fiber transmission.

\begin{figure}[!t]
  \centering
  \includegraphics[width=0.8\linewidth]{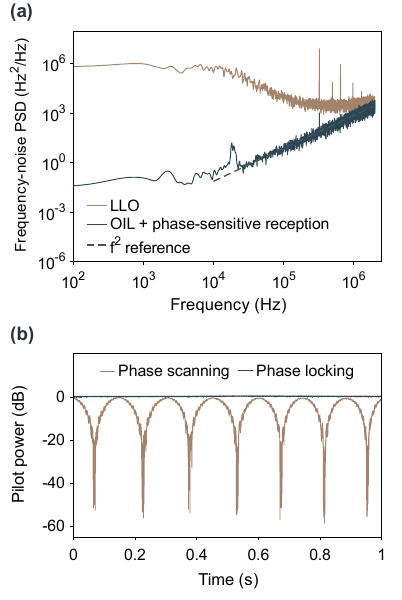}
  \caption{
  \textbf{Noise characterization and phase control of the receiver.}
  (a) Frequency-noise power spectral densities under free-running LLO operation without injection and OIL-assisted reception with the PLL closed.
  The dashed line indicates the \(f^{2}\) dependence associated with white phase noise.
  (b) Relative pilot power during LO phase scanning and phase locking.
}
  \label{figure3}
\end{figure}

\begin{figure*}[!t]
  \centering
  \includegraphics[width=0.9\linewidth]{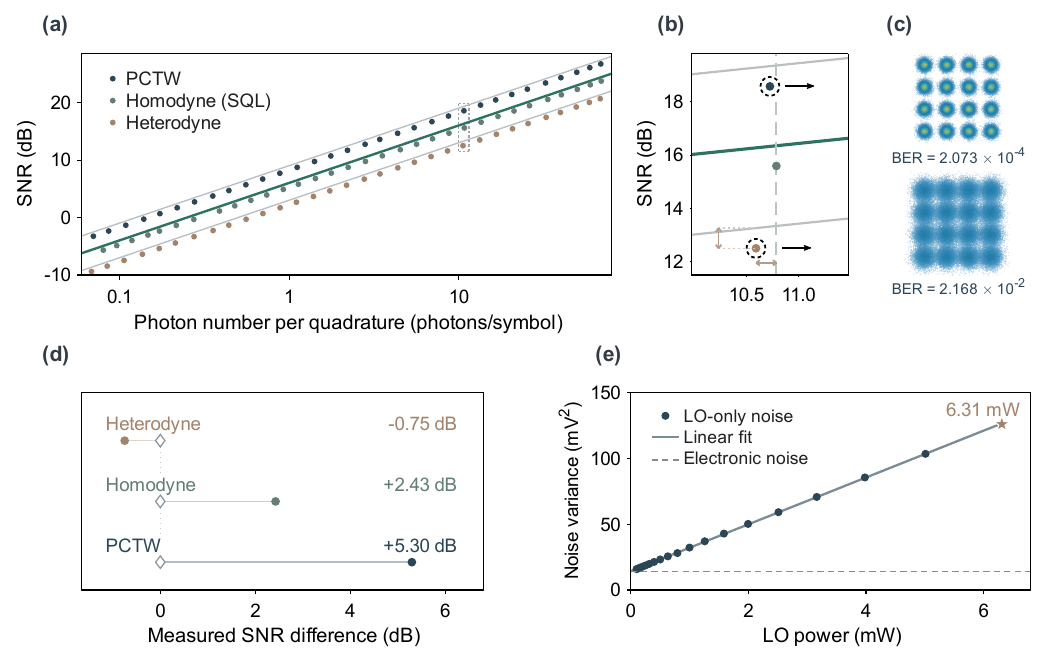}
  \caption{
  \textbf{Receiver performance and noise calibration for PCTW transmission over \(100~\mathrm{km}\).}
  (a) Measured SNRs (circles) and ideal predictions (solid lines) for PCTW, homodyne, and heterodyne reception at equal received signal-photon numbers per quadrature of each signal mode.
  The solid curve labeled ``Homodyne (SQL)'' denotes the ideal single-mode homodyne SQL benchmark.
  (b) Magnified view of the marked region in (a), showing results from a single acquisition.
  (c) PCTW (top) and heterodyne (bottom) 16-QAM constellations recovered from the acquisition in (b), with BERs shown below.
  (d) Mean SNR offsets over 32 acquisition settings.
  For each acquisition, the common reference is the ideal heterodyne SNR evaluated at the photon number per quadrature estimated for heterodyne reception.
  Filled circles indicate mean offsets, and open diamonds mark zero.
  (e) LO-only noise variance versus LO power, with solid and dashed lines indicating the linear fit and electronic-noise baseline, respectively.
  The star marks the \(6.31~\mathrm{mW}\) (\(8~\mathrm{dBm}\)) LO operating point.
}
  \label{figure4}
\end{figure*}

\begin{figure*}[!t]
  \centering
  \includegraphics[width=0.8\textwidth]{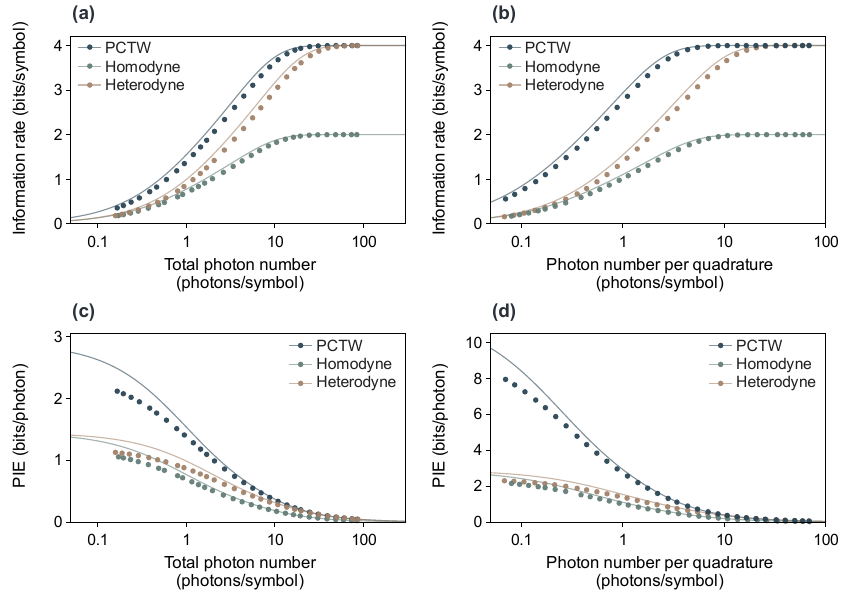}
  \caption{
  \textbf{16-QAM information rate and photon efficiency under two photon allocations.}
  (a) Achievable information rate versus total received signal-photon number.
  (b) Achievable information rate versus received signal-photon number per quadrature.
  (c) Photon information efficiency (PIE), defined as the information rate divided by the total received signal-photon number.
  (d) PIE defined as the information rate divided by the received signal-photon number per quadrature, in bits/photon.
  Solid lines show ideal theoretical predictions, and circles show values inferred from measured SNRs.
}
  \label{figure5}
\end{figure*}

\section{Experimental setup}

The experimental setup is shown in Fig.~\ref{figure2}.
The output of a \(1550~\mathrm{nm}\) laser with a linewidth below \(200~\mathrm{Hz}\) was split into signal and reference arms.
After IQ modulation, the signal was polarization-multiplexed with the reference and transmitted through \(100~\mathrm{km}\) of low-loss single-mode fiber with a total loss of \(16.5~\mathrm{dB}\).
At the receiver, the fields were polarization-demultiplexed, and the \(-40~\mathrm{dBm}\) reference injection-locked a slave laser whose \(8~\mathrm{dBm}\) output served as the regenerated LO.
The signal and LO were mixed and coherently detected by a low-noise balanced photodetector with an electrical bandwidth of \(1.6~\mathrm{GHz}\).

A 12-bit arbitrary waveform generator operating at \(12~\mathrm{GSa/s}\) produced three \(200~\mathrm{MBaud}\) test channels: homodyne at baseband, PCTW conjugate sidebands at \(\pm600~\mathrm{MHz}\), and heterodyne at \(1.2~\mathrm{GHz}\).
The transmitter digitally adjusted the channel amplitude weights to implement two allocations: equal signal photons per quadrature and equal total signal photons.
For the equal-total-photon measurements, the homodyne reference used balanced modulation of both quadratures, while only one quadrature was detected.
The three channels were transmitted synchronously within the same waveform and recorded in a single acquisition.

To maintain the phase-sensitive operating point, a \(2.5~\mathrm{MHz}\) auxiliary pilot was applied to the optical quadrature orthogonal to the data-bearing quadrature.
The detector output was split into feedback and measurement branches.
In the feedback branch, a lock-in amplifier extracted the phase-error signal for a proportional-integral-derivative (PID) controller, which drove a piezoelectric fiber stretcher in the LO path to compensate residual optical phase drift.
The measurement branch was digitized by a 10-bit oscilloscope at \(20~\mathrm{GSa/s}\).
Offline processing comprised digital downconversion, matched filtering, timing recovery and constellation phase correction.

This feedback requires a clearly resolvable phase-error signal extracted from the auxiliary pilot.
Under the \(100~\mathrm{km}\) transmission conditions of this experiment, the extracted phase-error signal was buried in noise when the uninjected slave laser operated as a free-running local oscillator (LLO), preventing reliable locking through pilot feedback alone.
Injection locking enabled the regenerated LO to track the reference derived from the same source and transmitted through the same fiber, thereby canceling most channel-induced phase fluctuations shared by the signal and reference during coherent detection.
The auxiliary-pilot phase-locked loop (PLL) further compensated residual low-frequency phase drift, completing the phase stabilization required for coherent combining of the conjugate twin waves.

Figure~\ref{figure3}(a) compares the frequency-noise power spectral densities under free-running LLO operation and with reference injection and the PLL enabled.
With the combined stabilization, low-frequency noise was reduced by several orders of magnitude, while the higher-frequency residual spectrum approached the \(f^{2}\) reference trend associated with white phase noise.

Figure~\ref{figure3}(b) shows the monitored pilot power during LO phase scanning and under phase locking.
During scanning, the pilot power exhibits periodic maxima and minima.
Under phase locking, it remains near the maxima of the scanning trace, indicating that closed-loop control maintains the signal--LO relative phase at the operating point of strongest coherent combining of the conjugate twin waves.

\section{Results}

\subsection{Receiver performance and noise calibration}

With the signal--LO phase stabilized, we compared the SNRs and 16-QAM signal recovery of the three receivers.
Figure~\ref{figure4}(e) characterizes the LO-dependent receiver noise.
With the signal blocked, the measured noise variance increased approximately linearly with LO power, consistent with the power dependence of shot noise.
The star marks the \(6.31~\mathrm{mW}\) (\(8~\mathrm{dBm}\)) LO operating point.

Figure~\ref{figure4}(a) shows the SNRs of the three receivers versus received signal-photon number per quadrature.
The solid curve labeled ``Homodyne (SQL)'' represents the ideal single-mode homodyne SQL benchmark, whereas the homodyne circles show measured receiver performance.
Under the stated per-mode photon normalization, the measured PCTW SNR exceeds this ideal benchmark across the plotted measurement range.
At equal signal photons per quadrature, the measured PCTW SNR exceeded the measured homodyne and heterodyne SNRs by approximately \(3~\mathrm{dB}\) and \(6~\mathrm{dB}\), respectively.
Figure~\ref{figure4}(b) shows a magnified view of results from a single acquisition.
Vertical offsets reflect SNR losses caused by residual system noise, whereas horizontal offsets reflect differences in photon-number estimates between channels during acquisition.
Figure~\ref{figure4}(d) summarizes the mean SNR offsets over 32 acquisition settings relative to a common ideal heterodyne reference.
For each acquisition, this reference is evaluated at the photon number per quadrature estimated for heterodyne reception.
The mean offsets are \(-0.75\), \(+2.43\), and \(+5.30~\mathrm{dB}\) for heterodyne, homodyne, and PCTW reception, respectively.

Figure~\ref{figure4}(c) shows 16-QAM constellations recovered from the same acquisition.
Compared with heterodyne reception, PCTW produced more compact and clearly separated symbol clusters, with the BER decreasing from \(2.168\times10^{-2}\) to \(2.073\times10^{-4}\), demonstrating that its SNR advantage translates into more reliable symbol decisions.

\begin{figure*}[!t]
  \centering
  \includegraphics[width=0.9\textwidth]{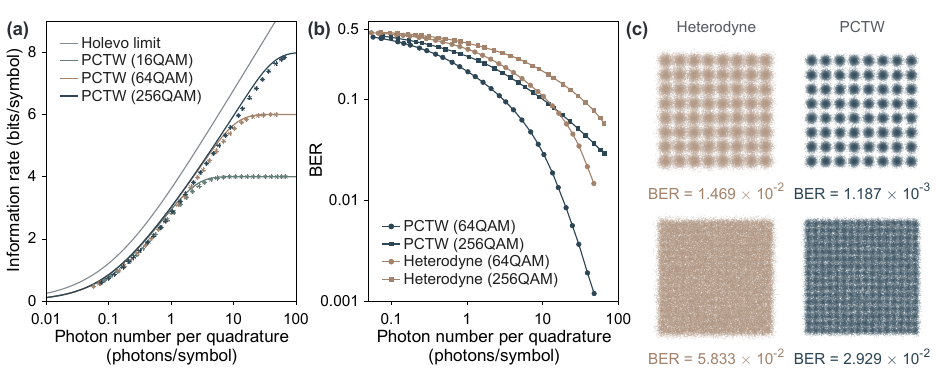}
  \caption{
  \textbf{Higher-order QAM performance of PCTW and heterodyne reception.}
  (a) PCTW information rates for 16-QAM, 64-QAM, and 256-QAM versus received signal-photon number per quadrature.
  Colored solid lines show theoretical rates, and stars show rates inferred from measured SNRs.
  The gray curve is the single-mode Holevo reference evaluated using the total received PCTW signal-photon number, which is four times the photon number per quadrature.
  (b) BERs of 64-QAM and 256-QAM under PCTW and heterodyne reception versus received signal-photon number per quadrature.
  (c) Recovered 64-QAM (top) and 256-QAM (bottom) constellations for heterodyne (left) and PCTW (right) reception, with BERs shown below.
}
  \label{figure6}
\end{figure*}

\subsection{Information rate and photon efficiency}

To assess the information-transmission benefits of the measured SNR advantage, we used the preceding 16-QAM model to obtain achievable information rates and photon-normalized efficiencies from the measured SNRs, as shown in Fig.~\ref{figure5}. 
The experimentally inferred results follow the trends of the ideal curves, with the shortfall before saturation reflecting the measured SNR penalties reported above.

For the balanced two-quadrature alphabet considered here, the ideal model predicts a higher information rate for PCTW than for homodyne and heterodyne reception before saturation at equal total received signal-photon numbers [Fig.~\ref{figure5}(a)].
Homodyne reception detects only one quadrature of 16-QAM, limiting its information rate to \(2~\mathrm{bits/symbol}\).
Heterodyne reception recovers both quadratures, allowing an information rate of up to \(4~\mathrm{bits/symbol}\).
PCTW also recovers the full 16-QAM symbol information, while coherent combining of the conjugate twin waves enables it to approach this limit with fewer total photons than heterodyne reception.

Figure~\ref{figure5}(c) compares the PIE under the same total photon budget. 
Under this modulation constraint, the ideal PCTW PIE exceeds those of homodyne and heterodyne reception before saturation.
As the photon number increases, the information rate grows more slowly than the photon number, causing the PIE to decrease for all three receivers. 
Once PCTW and heterodyne reception both approach \(4~\mathrm{bits/symbol}\), their PIE difference narrows.
The photon-utilization advantage of PCTW is thus most apparent before the finite-constellation information-rate limit is reached.

With equal photon numbers per quadrature, the information-rate advantage of PCTW is more pronounced at low and intermediate photon numbers [Fig.~\ref{figure5}(b)].
Compared with heterodyne reception, PCTW requires fewer photons per quadrature to achieve a given information rate, although both remain subject to the same 16-QAM information-rate limit.
Figure~\ref{figure5}(d) shows the PIE defined using the received signal-photon number per quadrature, for which PCTW likewise provides a higher normalized information yield before saturation.
This quantity is expressed in bits/photon.
Because the receivers use different total signal-photon numbers under this allocation, its denominator differs from that of the PIE in Fig.~\ref{figure5}(c), which is defined using the total received signal-photon number.

\subsection{Higher-order QAM performance}

To extend the information rate beyond the 16-QAM limit in Fig.~\ref{figure5}, we further investigated PCTW reception with 64-QAM and 256-QAM.
Figure~\ref{figure6}(a) compares the information rates of the three modulation formats as functions of the received signal-photon number per quadrature.
At low photon numbers, their information rates are similar, as reception is primarily limited by noise and increasing the constellation size provides little benefit.
As the photon number increases, 16-QAM approaches saturation, whereas the information rates of 64-QAM and 256-QAM continue to increase toward \(6\) and \(8~\mathrm{bits/symbol}\), respectively.
Higher-order modulation therefore extends the photon-number range over which increasing SNR translates into a higher information rate.
The rates inferred from the measured SNRs follow the corresponding theoretical trends.
At intermediate and high photon numbers, higher-order modulation also reduces the gap to the plotted single-mode Holevo benchmark relative to 16-QAM.
This benchmark is evaluated using the total received signal-photon number of PCTW~\cite{Giovannetti2004LossyCapacity}.

Higher modulation orders also make symbol decisions more demanding: at the same average symbol energy, denser constellations have smaller nearest-neighbor separations and are more sensitive to noise.
Figure~\ref{figure6}(b) compares the BERs of PCTW and heterodyne reception for 64-QAM and 256-QAM under equal received signal-photon numbers per quadrature.
PCTW achieves a lower BER for both modulation formats.
Over the BER range covered by both receivers, PCTW also requires fewer photons per quadrature to achieve a given BER.
Meanwhile, the BER of 256-QAM is generally higher than that of 64-QAM, indicating that the information-rate benefits of higher-order modulation require a higher received SNR.
The coherent-combining gain of PCTW helps meet these increased SNR requirements.

The recovered constellations in Fig.~\ref{figure6}(c) provide a visual comparison of this improvement.
For 64-QAM, neighboring symbol clusters overlap substantially under heterodyne reception, whereas PCTW produces tighter and more clearly separated clusters.
For the examples shown, the BER decreases from \(1.469\times10^{-2}\) for heterodyne reception to \(1.187\times10^{-3}\) for PCTW, a reduction by a factor of approximately \(12.4\).
For the denser 256-QAM constellation, PCTW also improves symbol distinguishability, although substantial cluster overlap remains.
The corresponding BER decreases from \(5.833\times10^{-2}\) to \(2.929\times10^{-2}\), approximately halving the error rate.
Both modulation formats show improved reception quality, with the magnitude of the BER improvement depending on the modulation order and the received SNR at which the comparison is made.

\subsection{Long-term stability}

The preceding results demonstrated that PCTW reception supports higher-order modulation formats, including 64-QAM and 256-QAM, while improving reception performance.
For long-distance transmission, sustaining these advantages over extended operation is equally important.
We therefore performed an extended reception test with 256-QAM signals transmitted over \(100~\mathrm{km}\) of fiber to assess the precise and sustained control of the signal--LO relative phase provided by the combined injection-locking and PLL scheme.

\begin{figure}[!t]
  \centering
  \includegraphics[width=0.9\linewidth]{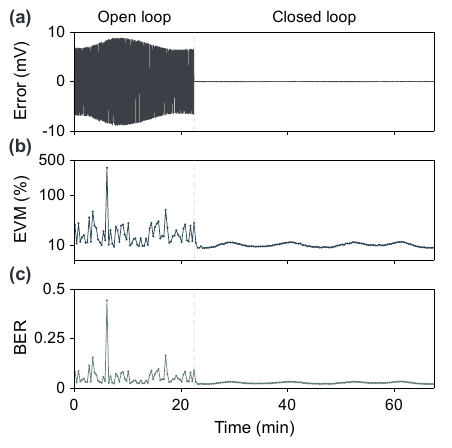}
  \caption{
  \textbf{Long-term stability of PCTW reception.}
  (a) Phase-error signal, (b) error vector magnitude (EVM), and (c) BER for 256-QAM PCTW reception over \(100~\mathrm{km}\) of fiber under open- and closed-loop PLL operation.
  Dashed lines mark the inferred boundary between the two loop states.
}
  \label{figure7}
\end{figure}

Figures~\ref{figure7}(a)--(c) show the phase-error signal, EVM and BER, respectively, under open- and closed-loop PLL operation.
Under open-loop operation, the error signal fluctuates markedly, and EVM and BER also exhibit large variations and occasional degradation.
Under closed-loop operation, the error signal remains near zero, while fluctuations in EVM and BER are substantially reduced, with both metrics remaining stable during approximately \(45~\mathrm{min}\) of closed-loop observation.
These results demonstrate that stable injection locking and precise closed-loop PLL control can maintain the relative phase between the signal and LO after \(100~\mathrm{km}\) of transmission, sustaining phase-sensitive reception performance for higher-order 256-QAM over extended operation.

\section{DISCUSSION AND CONCLUSION}

\subsection{Discussion}

Beyond the present demonstration, PCTW encoding and joint detection provide a framework for further advances in coherent reception through low-noise phase-sensitive amplification (PSA) and the incorporation of quantum-correlated optical states.

At the system level, low-noise parametric optical amplification is a natural extension towards longer transmission distances~\cite{Caves1982AmplifierNoise,Olsson2018LongHaulPSA,Kakarla2020OnePhoton}.
The signal and phase-conjugate modes in PCTW can serve as the signal and idler inputs of a non-degenerate phase-sensitive amplifier~\cite{Andrekson2020PSAReview}.
With appropriate phase matching and stabilization of the relative pump--PCTW phase, the PSA can amplify the joint quadratures selected by the PCTW receiver and reduce the impact of subsequent detection loss and electronic noise~\cite{Shaked2018ParametricHomodyne}.
Low-noise PSA in monolithic nonlinear waveguides offers another route toward compact integration with the PCTW receiver~\cite{Ye2021MonolithicPSA}.

The two modes of the PCTW scheme considered here are correlated through their classically imposed modulation, while their vacuum fluctuations remain independent for a given encoded symbol.
A quantum-enhanced extension would apply the phase-conjugated modulation to a two-mode squeezed state with correlations matched to the measured observables \(\widehat U\) and \(\widehat W\)~\cite{SchafermeierAndersen2018, Schnabel2017SqueezedLight} such that the noise variances of both observables could fall below the vacuum-noise level and the signal SNR could be improved further.
Related experiments have demonstrated several ingredients relevant to this quantum extension.
Squeezed-light-enhanced heterodyne detection has achieved simultaneous noise reduction in the signal and image bands~\cite{Anai2024Heterodyne}.
Entanglement-assisted joint-quadrature communication has also been demonstrated over a 20~km fiber channel~\cite{Ren2026FiberDenseCoding}, while we have demonstrated remote preparation of bright squeezed light using field-deployed fibers~\cite{Wang2025RemoteSqueezed}.
Building on these developments, a quantum-correlated PCTW link would require preserving joint squeezing through propagation and matching its orientation to the receiver measurement.

\subsection{Conclusion}

We have experimentally demonstrated injection-locking-assisted PCTW reception of 200-MBaud signals with modulation formats up to 256-QAM over 100~km of optical fiber.
By combining optical injection locking for LO regeneration with auxiliary-pilot optical phase feedback, the receiver maintains the phase relation required for coherent combining of the conjugate modes using a single balanced photodetector.
At equal received signal-photon number per quadrature in each data-carrying optical mode, the measured PCTW SNR exceeds the ideal single-mode homodyne SQL benchmark.
The SNR gains over the measured homodyne and conventional heterodyne results are approximately 3~dB and 6~dB, respectively.
Constellation and BER measurements confirm that this gain improves the recovery of higher-order QAM signals.
Complementary information-rate and PIE analyses quantify the benefits under explicitly defined signal-photon budgets.
These results advance phase-conjugate joint reception to phase-stabilized, long-distance transmission of high-order complex modulation and establish an experimentally validated receiver architecture for developing photon-efficient coherent optical links.

\begin{backmatter}

\bmsection{Funding}
This work was supported in part by the National Natural Science Foundation of China under Grant 62475297, in part by the Guangdong Basic and Applied Basic Research Foundation under Grant 2024B1515020004, and in part by the International Science and Technology Cooperation Program of Guangdong Province under Grant 2026A0505020042.

\bmsection{Disclosures}
The authors declare no conflicts of interest.

\bmsection{Data Availability Statement}
Data underlying the results presented in this paper are not publicly available at this time but may be obtained from the authors upon reasonable request.

\end{backmatter}

\bibliography{references}

\end{document}